\documentstyle[12pt]{article}
\begin{document}
\bibliographystyle{plain}

\date{RAL-93-011\\23 March 1993}

\title{ Probes of heavy meson substructure in $e^+e^-$ annihilation}

\author{F.E.Close and G.J.Gounaris\footnotemark
\\
Rutherford Appleton Laboratory, \\
Chilton, Didcot,Oxon OX11 0QX, England}
\maketitle

\begin{abstract}

We apply Heavy Quark Effective Theory to the production of
$0^-$ and $1^-$ $Q\bar{q}$ states in $e^+ e^-$ annihilation. We show that HQET
implies that the electric quadrupole amplitudes vanish and we propose tests
for this theory.
We also show how HQET can be applied to distinguish the
$^3D_1$ and $^3S_1$ $Q\bar{Q}$ states.

\end{abstract}
\centerline{(submitted to Physics Letters B)}
\
\footnotetext{Permanent address:
Dept of Theoretical Physics, University of Thessaloniki, Greece. Partially
supported by C.E.C. project SC1-CT91-0729 }

\newpage





Heavy Quark Effective Theory (HQET)[1]
 has been extensively investigated for the cases where a heavy quark
undergoes a (flavour changing) current induced transition (such as
$B \rightarrow D^*l\nu$). This has been
most widely applied where $y=v\cdot v' \approx 1$ with $v,v'$
the four-velocities of the initial and final hadrons (heavy quarks). There is
another
physical region where heavy quark interactions with (electromagnetic) currents
are important, namely heavy flavour pair production in electron-positron
annihilation (such as at a $\tau$-Charm or B-factory). Some old
results[2,3]
have recently been reformulated within the context of HQET for such processes.
It is the purpose of this note to advertise how $e^+e^-$ annihilation
may be analysed
in order to test the HQET. Furthermore we shall show how the production of
bottom, charm and even
{\it strange} particles may be of interest and how these ideas may be exploited
to determine the structure of the $\psi$, or $\Upsilon$, resonance states.

In a quark model analysis where the pseudoscalar and
vector $Q\bar{q}$ are assumed to be the $^1S_0$ and $^3S_1$ members of an
$SU(2)_{spin}$ supermultiplet, it is known that
the production of such states in $e^+ e^-$
annihilation involves only three a priori arbitrary form factors, in contrast
to the most general case where five independent form factors are
needed. Denoting
these as the $F_E$ (Electric), $F_M$ (Magnetic) and $F_Q$
(Quadrupole), then the relative production cross sections are[3]

\begin{eqnarray}
\sigma (e^+e^- \rightarrow PP): \sigma (e^+e^- \rightarrow PV + VP):
 \sigma (e^+e^- \rightarrow VV) & = & \nonumber \\
1 : 4\frac{s}{4M^2}(\frac{F_M}{F_E})^2 : 3+ 4\frac{s}{4M^2} (\frac{F_M}{F_E})^2
 +
\frac{8}{9}
(\frac{s}{M^2})^2(\frac{F_Q}{F_E})^2
\end{eqnarray}

\noindent where $V \equiv B^*,D^*,$ and $P \equiv B,D,$ are $Q\bar{q}$
mesons made of heavy-light
quarks. In eq(1), (as well as in the rest of this paper), it has been assumed
that the $e^+e^-$ energy is sufficiently above threshold that the $V$ and $P$
mesons can be considered to be degenerate. In this approximation the heavy
quark
mass $M$ is taken to be equal to that of the heavy meson.

  In HQET these channels are described by a single form factor.
 In the language of ref [3]
this means that $F_E = F_M$ and
$F_Q=0$. The significance of these
constraints was noted soon after the discovery of charm[2,3]
but the HQET has recently put them on a sounder footing [4,5].

Setting $F_Q=0$ in the general formula, eq(1),
leads to a sum rule for the production
differential cross sections at any angle $\theta$ to the initial $e^+e^-$
axis,

\begin{equation}
3\frac{d\sigma}{d\theta} (e^+e^- \rightarrow PP) +
\frac{d\sigma}{d\theta}(e^+e^- \rightarrow PV + VP)
 = \frac{d\sigma}{d\theta}(e^+e^- \rightarrow VV)
\end{equation}

 In HQET one expects that

\begin{equation}
V = ^3S_1 + O(\frac {1}{M^2})^3D_1
\end{equation}

\noindent This implies that as the heavy quark mass $M \rightarrow \infty$,
the vector meson $V \rightarrow ^3S_1$ which in turn means
that $F_Q \rightarrow 0$.
So eq(2) may be interpreted as a test of HQET at leading order, and in
particular as a test of the $^3S_1$ nature of the $Q\bar{q}$ vector.

 The individual contributions of the various final states to the sum rule
eq(2) depend upon the dynamic coupling
of the {\it initial} $e^+e^-$ to the heavy quarks (e.g. whether in the
continuum or on an S or D-wave $Q\bar{Q}$ resonance). So by varying the
beam energy we can expect different ratios of the individual contributions
to eq(2).

In the continuum
 the direct coupling of the photon to the $Q\bar{Q}$ pair
involves a $\gamma _\mu$ vertex together with perturbative QCD corrections
which induce also a
$\sigma _{\mu \nu} q^{\nu}$ form factor. In this case the ratios of the
total cross sections are[4,5]

\begin{equation}
\sigma(e^+e^- \rightarrow PP:PV+VP:VV) = 1+h :4 \frac{s}{4M^2} : 3(1+h) +
4\frac {s}{4M^2}
\end{equation}

\noindent where
\begin{equation}
h \equiv - \frac{2 \alpha_s}{3 \pi} \sqrt{1-\frac{4M^2}{s}}
log\{\frac{s}{2M^2} -1 +\frac{s}{2M^2}\sqrt{1-\frac{4M^2}{s}}\}
\end{equation}

\noindent describes the first order
QCD corrections[5].
 Note that eq(4) gives a particular realisation
of eq(2), which
is a consequence of the fact that the off shell photon has no electric
quadrupole coupling to the VV final state.

On an S-wave (or D-wave) resonance we explicitly neglect D-wave (or S-wave)
 contributions respectively. Thus although we cannot predict the absolute
magnitudes of the
form factors, their relative strengths follow simply from angular momentum
considerations alone and are independent of perturbative QCD corrections
at the heavy quark production vertex.
For an S-wave ($^3S_1(Q\bar{Q})$) bound state
 we find

 \begin{eqnarray}
 <P(v_1)\bar{P}(v_2)|^3S_1, \epsilon>  =  M\frac{(1+2v_1^0)}{3}\xi (v_1 \cdot
v_
 2)
(v_1 - v_2)_\mu \cdot \epsilon ^\mu  \\
 <V(v_1,\epsilon_1)\bar{P}(v_2)|^3S_1,\epsilon>  =  iM\frac{(1+2v_1^0)}{3v_1^0}
\xi (v_1 \cdot v_2)\epsilon _{\mu \nu \lambda \sigma}\epsilon ^\mu
\epsilon_1 ^{* \nu} v_1^\lambda v_2^\sigma  \\
 <V(v_1, \epsilon_1)\bar{V}(v_2, \epsilon_2)|^3S_1, \epsilon>  =
M \xi (v_1 \cdot v_2) \frac {(1+2v_1^0)}{3(1+v_1^0)} \times \nonumber \\
\{ (1+v_1^0)(\epsilon _1^*
\cdot \epsilon _2^*) (v_1 - v_2)_{\mu}
- (1+1/v_1^0)[(\epsilon _2^* \cdot v_1)
\epsilon _{1 \mu}^* - (\epsilon _1^* \cdot v_2)
\epsilon _{2 \mu}^*] \nonumber \\
 - \frac{1}{2v_1^0} (\epsilon _1^* \cdot v_2)
(\epsilon _2^* \cdot v_1)(v_1 - v_2)_\mu \} \epsilon ^{\mu}
 \end{eqnarray}

\noindent where $\epsilon$ is the polarisation vector of the decaying state,
$v_ 1^0 =
\frac{\sqrt{s}}{2M}$ and $v_1,v_2$ are the four-velocities
 of the mesons in the final state. This
leads to the following realization of eq(2) for the integrated cross sections

\begin{equation}
\sigma(e^+e^- \rightarrow PP:PV+VP:VV) = 1: 4 :7
\end{equation}

\noindent whereas for a D-wave ($^3D_1(Q\bar{Q})$) bound state one finds

\begin{eqnarray}
 <P(v_1)\bar{P}(v_2)|^3D_1,\epsilon>  =  -M\frac{2(v_1^0-1)}{3}\xi (v_1 \cdot
v_
 2)
(v_1 - v_2)_\mu \cdot \epsilon ^\mu \\
 <V(v_1,\epsilon_1)\bar{P}(v_2)|^3D_1,\epsilon>  =  iM\frac{(v_1^0-1)}{3v_1^0}
\xi (v_1 \cdot v_2)\epsilon _{\mu \nu \lambda \sigma}\epsilon ^\mu
\epsilon_1 ^{* \nu} v_1^\lambda v_2^\sigma  \\
 <V(v_1, \epsilon_1)\bar{V}(v_2, \epsilon_2)|^3D_1, \epsilon>  =
-M \xi (v_1 \cdot v_2) \times \nonumber \\
 \{\frac{2(v_1^0-1)}{3} (\epsilon _1^*
\cdot \epsilon _2^*) (v_1 - v_2)_{\mu}  + \frac{(v_1^0-1)}{3v_1^0}
[(\epsilon _2^* \cdot v_1)
\epsilon _{1 \mu}^* - (\epsilon _1^* \cdot v_2)
\epsilon _{2 \mu}^*] \nonumber \\
- \frac{1+2v_1^0}{6v_1^0(1+v_1^0)}
 (\epsilon _1^* \cdot v_2)
(\epsilon _2^* \cdot v_1)(v_1 - v_2)_\mu \} \epsilon ^{\mu}
\end{eqnarray}

\noindent which leads to

\begin{equation}
\sigma(e^+e^- \rightarrow PP:PV+VP:VV) = 1: 1 :4
\end{equation}

\noindent (see also[3,6])


Note that the continuum result differs from the S-wave only in the
perturbative QCD corrections which vanish at threshold and are small in the
kinematic region
 of interest. On the other hand the D-wave result given in eq(13) is very
different from both eq(4) and eq(9) for the continuum and S-wave cases
respectively. This is related to the fact that the D-wave contribution also
vanishes at threshold like $\vec{v}^2$, which can be seen by inspection of
eqs(10-12). The D-wave result in eq(13) naturally gives the dominant
contributio
 n
on a D-wave resonance, provided that this resonance lies sufficiently above
 threshold to justify the neglect of the mass difference between the $P$
 and $V$ states.
Thus to the extent that eq(2) is realised in
the data, we can use eq(13) to identify the $^3D_1 \psi$ and $\Upsilon$-like
 states.
At this point it is worthwhile to emphasise that within the framework of HQET
we can determine the internal structure of the $Q\bar{Q}$ resonance by studying
only the branching ratios into various channels without need for detailed
angular distributions.

These results suggest the
following strategy.

\vskip 0.1in

{\it Possibility 1}

\vskip 0.1in

Eq(2) is violated or, in the continuum, eq(4) is violated.
 In this case HQET at leading order in $M_Q$
is not a good approximation.

One particular source of such violation could be the presence of $F_Q \neq 0$.
This can be tested by analysing the polarisation of the final state vector
 mesons[3,7].
According to our treatment, in
 this case the vector meson wavefunction should in general not be
simply given by $^3S_1$; this
would undermine some of the analysis of semileptonic decays of heavy flavours
such as $B \rightarrow D^* l \nu$ (which, in HQET, implicitly
assumes no $^3D_1$ component in the $D^*$).
It may even be interesting to study $e^+e^- \rightarrow K^*\bar{K^*}$,
e.g. at $DA\Phi NE$, where it would be natural to expect a non-vanishing
$F_Q$, (with corresponding implications for HQET applications to
$B,D \rightarrow K^*l\nu$).Independent of the specific interest in HQET, this
is a unique
way of measuring the quadrupole moments of vector mesons.

\vskip 0.1in

{\it Possibility 2}

\vskip 0.1in

Eq(2) and (in the continuum) eq(4) are satisfied, and $F_Q = 0$.

 In this case HQET is correct to a high
 level of accuracy and the vector state $V(Q\bar{q}) = ^3S_1$.Such a
direct measure of the vector meson's wavefunction would constrain models
of substructure; in particular it would eliminate the possible presence of
a significant mass-independent $\sigma \cdot \sigma$ interaction
between the quarks as would arise from an elementary
 $\gamma _5$ pseudoscalar interaction.
In such a case it will also be interesting to study the predictions of
this theory not only for bottom and charmed states but also for production
of strange mesons in order to help
establish the extent to which the strange quark
may be considered as  heavy. If $e^+e^- \rightarrow K^*\bar{K^*}$ shows that
$F_Q \neq 0$, then some applications of leading order HQET to $B(D)
\rightarrow K^* l \nu$ or $b \rightarrow s \gamma$ may need reexamination.

If HQET is established in the continuum, then measurement of the production
ratios on $Q\bar{Q}$ resonance may, courtesy of eqs (9, 13), be able to
determine the $^3S_1, ^3D_1$ content of the $Q\bar{Q}$ states.

The essential physics that enabled this analysis is that the initial
state has  well defined total angular momentum and parity quantum numbers,
 which in this case
happen to be $1^-$, with the fermions in either S or D orbital angular
momentum. Such
ideas could also be
applied to final states consisting of spin-$\frac{1}{2}, \frac{3}{2}$ baryons
in $e^+e^-$ annihilation and to
$\gamma \gamma$ physics where the $\vec{L}$ of the various $\chi$ states could
be
determined. For example, when $J^{PC} = 2^{++}$ the $\chi$ can be either
$^3P_2$ or $^3F_2$. Ref.[8]
 has shown how the helicity dependence of
$\gamma \gamma$ production is sensitive to this. These are particular examples
of a more general possibility, namely that the $\vec{L}$ and $\vec{S}$
substructure of a heavy state of total $\vec{J}$ may be analysed by just
the branching ratios to various channels.

To recapitulate: In $e^+e^- \rightarrow PP,PV,VV$ we have identified
ways of determining
how good HQET is to leading order in $M_Q$. If HQET turns out to be valid,
such processes can then be used to study the substructure of heavy states.

{\large\bf Acknowledgements}

G.J.G would like to thank Rutherford Appleton Laboratory and its Theory
Group for the hospitality enjoyed during the preparation of the present work.

\newpage

{\large\bf References}
\vskip 0.1in
[1] N. Isgur and M.B. Wise, Phys.\ Lett.\ B {\bf 232}, 113 (1989);
{\bf 237}, 527 (1990).
\vskip 0.1in
[2] A.De Rujula, H.Georgi and S.Glashow, Phys.Rev.Letters {\bf 37},
398 (1976)
\vskip 0.1in
[3] F.E.Close, Phys.Lett.B.{\bf 65}, 55 (1976)

    F.E.Close and W.N.Cottingham, Nucl. Phys.B.{\bf 99}, 61 (1975)
\vskip 0.1in
[4] T. Mannel, W. Roberts, and Z. Ryzak, Nucl.\ Phys.\ {\bf B368}, 204
(1992).
\vskip 0.1in
[5] A.Falk and B.Grinstein, Phys. Lett.B.{\bf 249},314 (1990)

\vskip 0.1in

[6] S.Matsuda, Phys. Lett. B {\bf 66}, 70, (1977)
\vskip 0.1in

[7] P.Bialas, F.E.Close, J.Koerner and K. Zalewski, RAL-93-012
(unpublished)
\vskip 0.1in
[8] E.Ackleh, T.Barnes and F.E.Close, Phys. Rev {\bf D46}, 2257 (1992);

    T.Barnes, F.E.Close and Z.P.Li, Phys. Rev {\bf D43}, 2377 (1990)


\end{document}